\documentclass[12pt,thmsa]{article}
\usepackage{amsfonts}

\usepackage{sw20rui}

\input{tcilatex}
\usepackage[T1]{fontenc}
\begin{document}

\author{Ezra T. Newman \\
%EndAName
Dept. of Physics and Astronomy, University \\
of Pittsburgh, Pgh., PA,15260, USA}
\title{Asymptotic Twistor Theory and the Kerr Theorem \\
PACS: 02.40.Tt, 04.20.-q, 04.20.Ha}
\date{12.13.05 }
\maketitle

\begin{abstract}
We first review asymptotic twistor theory with its real subspace of null
asymptotic twistors. This is followed by a description of an asymptotic
version of the Kerr theorem that produces regular asymptotically shear free
null geodesic congruences in arbitrary asymptotically flat Einstein or
Einstein-Maxwell spacetimes.
\end{abstract}

\section{Introduction}

This work is concerned with certain geometric structures which naturally
exist on Penrose's future null infinity, $\frak{I}^{+},$ that are associated
with any given asymptotically flat Einstein or Einstein-Maxwell space-time$.$
We take $\frak{I}^{+}$ to be $S^{2}xR$ and describe the structures in an
arbitrary but given Bondi coordinate system, ($u,\zeta ,\overline{\zeta }$).
We further assume that all the functions associated with the real physical
space-time are analytic functions that can be extended, at least, a short
way into the complexified space-time.

We review how, from solutions to the `good-cut equation'\cite
{H-space1,H-space2} for any asymptotically flat Einstein or Einstein-Maxwell
space-time, one defines the three-complex dimensional asymptotic projective
twistor space\cite{PM,RP,NPK,Hansen}, $\frak{T,}$ and its restriction to the
real five-dimensional subspace of null projective twistors P$\frak{N.}$ P$%
\frak{N,}$ which possesses a CR structure, can then be realized as null
infinity, $\frak{I}^{+},$ with the sphere of past null directions, $S^{2},$
attached to each point of $\frak{I}^{+},$ i.e., as the sphere bundle over $%
\frak{I}^{+}.$ By choosing an arbitrary holomorphic curve\cite{NKI} in the
solution space of the good cut equation, i.e., in H-space, an \textit{\
asymptotically} shear-free null geodesic congruence can be constructed.
Using this construction, we show that there is an asymptotic version of the
Kerr theorem. This version of the theorem says that a holomorphic surface in 
$\frak{T}$ that intersects with P$\frak{N}$ produces a \textit{regular
asymptotically} shear free null geodesic congruence in a space-time
neighborhood of $\frak{I}^{+}.$ By using this version of the Kerr theorem,
for each choice of the arbitrary holomorphic curve, a three-dimensional CR
structure is induced on $\frak{I}^{+}.$

This work can be considered as giving a generalization, to arbitrary
asymptotically flat Einstein-Maxwell space-times, of the work of Nurowski
and Trautman\cite{PNAT} on the Robinson-structures that are associated with
certain specific space-times. In addition it generalizes ideas of Penrose
and Rindler that are associated with flat twistor theory and it gives a
concrete description of Penrose's hypersurface twistor version of the Kerr
theorem\cite{PandR,PC} but now applied to $\frak{I}^{+}$.

\section{Asymptotic Twistor Theory}

We begin with an asymptotically flat, Einstein or Einstein-Maxwell
space-time $\frak{M}$ and its associated future null infinity, $\frak{I}
^{+}, $ with Bondi coordinates $(u$,$\zeta ,\overline{\zeta }$) and a given
(complex) Bondi shear

\begin{equation}
\sigma =\sigma (u,\zeta ,\overline{\zeta })  \label{sigma}
\end{equation}
which we assume is an analytic function in its three arguments. We then
consider the complexification of $\frak{I}^{+}$, to $\frak{CI}^{+},$ with
the three independent complex coordinates $(u$,$\zeta ,\widetilde{\zeta })$
with $\overline{\zeta }\Rightarrow \widetilde{\zeta },$ and the analytic
extension of the shear to $\sigma =\sigma (u,\zeta ,\widetilde{\zeta })$ in
a small neighborhood around the real $\frak{I}^{+}.$

The H-space associated with $\frak{M}$ is the four complex dimensional
solution space (with local coordinates $z^{a}$) to the so-called `good-cut
equation'\cite{H-space1,H-space2},

\begin{equation}
\overline{\text{\dh }}^{2}Z=\overline{\sigma }(Z,\widetilde{\zeta },\zeta ).
\label{good cut}
\end{equation}
The solutions, which can be written as 
\begin{equation}
u=Z(z^{a},\zeta ,\widetilde{\zeta }),  \label{solutions}
\end{equation}
define, for each fixed value of $z^{a},$ a four parameter set of complex
surfaces in $\frak{CI}^{+}.$ For the moment, rather than considering the
solutions as surfaces in $\frak{CI}^{+},$ we consider them at a \textit{%
fixed value} of $\zeta =\zeta _{0}$ and then interpret them as complex
curves in the complex $(u,\widetilde{\zeta })$ space that satisfy the second
order differential equation, Eq.(\ref{good cut}). The solutions are defined
by two initial conditions, the value of Z and its first $\widetilde{\zeta }$
derivative at some \textit{arbitrary initial value} of $\widetilde{\zeta }.$
We take this initial value to be the complex conjugate of the value of $%
\zeta _{0},$ i.e., we take Z and its first $\widetilde{\zeta }$ derivative
at $\widetilde{\zeta }=\overline{\zeta }_{0}$ as the $initial$ conditions.

The initial conditions then are written as 
\begin{eqnarray}
u_{0} &=&Z(z^{a},\zeta _{0},\overline{\zeta }_{0})  \label{u0} \\
\overline{L}_{0} &=&\overline{\text{\dh }}Z(z^{a},\zeta _{0},\overline{\zeta 
}_{0})=P_{0}\partial _{\overline{\zeta }_{0}}Z(z^{a},\zeta _{0},\overline{
\zeta }_{0})  \label{L0} \\
P &=&1+\zeta \widetilde{\zeta },\text{ }P_{0}=1+\zeta _{0}\overline{\zeta }%
_{0}
\end{eqnarray}

\begin{definition}
Penrose\cite{PandR} defines projective asymptotic twistor space, $P\frak{T,}$
(a three complex (six real) dimensional space) as the set of curves in $%
\frak{CI}^{+}$ obtained from the triple ($\zeta _{0},u_{0},\overline{L}_{0}$
) \textit{without the restriction} of the initial $\widetilde{\zeta }$ being
the complex conjugate of $\zeta _{0}.$ Arbitrary choices of $\widetilde{%
\zeta }$ simply lead to arbitrary choices of coordinates for the description
of $P\frak{T.}$ We chose this initial value, $\widetilde{\zeta }=\overline{%
\zeta }_{0},$ for its use in the definition that Penrose gives for \emph{null%
} asymptotic twistors\cite{PandR,NPK}.
\end{definition}

In order to define null projective twistors we point out that there are a
set of dual curves (projective dual asymptotic twistors) defined from
solutions of the complex conjugate good cut equation, namely from $\dh ^{2}%
\overline{Z}=\sigma (\overline{Z},\zeta ,\widetilde{\zeta }),$ with its own
set of initial conditions analogous to Eqs.(\ref{u0}) and (\ref{L0}).

\begin{remark}
One can convert the \textit{projective asymptotic twistors} to \textit{\
asymptotic twistors\cite{NPK} }by introducing a rescaling parameter. We will
not discuss this issue since it is not needed here.
\end{remark}

\begin{definition}
A \textit{null} projective asymptotic twistor is a curve given so that the
complex conjugate dual curve (given by complex conjugate initial conditions)
has a real value for $u_{0}$ at $\widetilde{\zeta }_{0}=\overline{\zeta }%
_{0}.$ We thus have, for a null twistor, that $u_{0}=\overline{u}_{0}$ with $%
\overline{L}_{0}=P_{0}\partial _{\overline{\zeta }_{0}}Z(z^{a},\zeta _{0},%
\overline{\zeta }_{0})$ at $\widetilde{\zeta }_{0}=\overline{\zeta }_{0}.$
The space of null projective asymptotic twistors is referred to as $P\frak{N.%
}$ This definition comes from the vanishing of a scalar product\cite{NPK} on
asymptotic twistor space whose details are also not needed here.
\end{definition}

In order to make these definitions of asymptotic twistors and asymptotic
dual twistors closer to the usual definitions, we define the asymptotic
projective twistor coordinates 
\begin{equation}
\mu ^{0}\text{ , }\mu ^{1},\text{ }\zeta  \label{DT}
\end{equation}
from 
\begin{eqnarray}
u_{0} &=&\frac{\mu ^{0}+\overline{\zeta }_{0}\mu ^{1}}{P_{0}}  \label{DT2} \\
\overline{L}_{0} &=&\frac{\mu ^{1}-\zeta _{0}\mu ^{0}}{P_{0}}  \nonumber \\
\zeta &=&\zeta _{0}  \nonumber
\end{eqnarray}
so that

\begin{eqnarray}
\mu ^{0} &=&u_{0}-\overline{L}_{0}\overline{\zeta }_{0}\text{ }  \label{DT3}
\\
\mu ^{1} &=&\overline{L}_{0}+\zeta _{0}u_{0}  \nonumber \\
\zeta &=&\zeta _{0}  \nonumber
\end{eqnarray}

\begin{remark}
The special case of flat space twistors arises when the Bondi shear
vanishes, i.e. when the good cut equation is 
\begin{equation}
\overline{\text{\dh }}^{2}Z=0  \label{flatGoodCut}
\end{equation}
with projective twistor curves given by the solutions, with fixed values ($%
\mu ^{0},\mu ^{1},\zeta $), 
\begin{equation}
u=Z=\frac{\mu ^{0}+\widetilde{\zeta }\mu ^{1}}{1+\zeta \widetilde{\zeta }}.
\label{flatSolution}
\end{equation}
For this flat case, note that both $\mu ^{0}\ $and $\mu ^{1}$ are \textit{%
independent} of $\widetilde{\zeta },$ so that we need not require that $%
\widetilde{\zeta }_{0}=\overline{\zeta }_{0}.$
\end{remark}

Since in all that follows we will use the restricted initial point, $%
\widetilde{\zeta }_{0}=\overline{\zeta }_{0},$ we can safely drop the use of
the subscript index `0' and simply write $\widetilde{\zeta }=\overline{\zeta 
}.$ Also since no use will made of the rescaling of the projective twistors
to twistors, we will drop the adjective `projective' .

Returning to Eq.(\ref{DT3}) and from the definition of null asymptotic
twistors, we can define the real five dimensional CR manifold, $P\frak{N,}$
(the real subspace of null twistors) in the following fashion. $P\frak{N}$
is explicitly given parametrically in terms of the five real parameters ($u,$
the real and imaginary parts of $L$ and $\zeta $) by

\begin{eqnarray}
\mu ^{0} &=&u-\overline{L}\overline{\zeta }\text{ }  \label{N1} \\
\overline{\mu }^{0} &=&u-L\zeta  \label{N2} \\
\mu ^{1} &=&\overline{L}+\zeta u  \label{N3} \\
\overline{\mu }^{1} &=&L+\overline{\zeta }u  \label{N4} \\
\zeta &=&x+iy  \label{N5} \\
\widetilde{\zeta } &=&\overline{\zeta }=x-iy.  \label{N6}
\end{eqnarray}

A representative element of the equivalence class that defines the CR
structure of $P\frak{N}$ is given by the exterior derivatives of Eq.(\ref{N1}%
),(\ref{N3}) and (\ref{N5}), 
\begin{eqnarray}
d\mu ^{0} &=&du-\overline{L}d\overline{\zeta }-\overline{\zeta }d\overline{L}
\label{CRS1} \\
d\mu ^{1} &=&d\overline{L}+\zeta du+ud\zeta  \nonumber \\
d\zeta &=&dx+idy  \nonumber
\end{eqnarray}
which, by manipulation within the equivalence class, can be converted to 
\begin{eqnarray}
l &=&du-\frac{L}{(1+\zeta \overline{\zeta })}d\zeta -\frac{\overline{L}}{
(1+\zeta \overline{\zeta })}d\overline{\zeta }  \label{CR1} \\
N &=&\frac{d\overline{L}}{\overline{L}}+\frac{\zeta }{(1+\zeta \overline{%
\zeta })}d\overline{\zeta }  \label{CR2} \\
\overline{m} &=&\frac{d\zeta }{(1+\zeta \overline{\zeta })}.  \label{CR3}
\end{eqnarray}
The form $l$, is real while $N$ and $\overline{m}$ are complex.

$P\frak{N}$ can thus be realized as the (real) three-dimensional $\frak{I}%
^{+}$, ($u,\zeta ,\overline{\zeta }$), with the two-sphere, ($L,\overline{L}$
), of past light-cone directions attached to each point of $\frak{I}^{+}$,
i.e., as the sphere bundle over real $\frak{I}^{+}$.

\section{Generalized Kerr Theorem}

A further question arises: Is there a generalization of the Kerr Theorem
from flat twistor theory to asymptotic twistor theory. For unity of
presentation we first describe, in our language, the flat-space Kerr Theorem%
\cite{PandR} followed by a special case, the \textit{restricted} flat-space
Kerr theorem. This restricted class produces regular families of
asymptotically shear-free null geodesics..

\begin{theorem}
Any analytic function on projective twistor space generates a shear-free
null geodesic congruence in Minkowski space, i.e., from $F(\mu ^{0},\mu
^{1},\zeta )\equiv F(u-\overline{L}\overline{\zeta },\overline{L}+\zeta
u,\zeta )=0,$ one can construct a shear-free null geodesic congruence in
Minkowski space.
\end{theorem}

Geometrically, this relationship defines a function $\overline{L}$ = $%
\overline{L}(u,\overline{\zeta },\zeta ),$ (a null direction field on $\frak{%
I}^{+})$ which when extended into the interior of Minkowski space forms a
shear-free null geodesic congruence$.$

We consider the following special or restricted case of the Kerr theorem for
our later generalization. Restricting ourselves to the regular solutions (on
real $\frak{I}^{+}$ ) of the flat-space good cut equation, Eq.(\ref
{flatGoodCut}), which are obtained by the restriction of the ($\mu ^{0},\mu
^{1}$) in (\ref{flatSolution}) to be linear functions of $\zeta ,$(See
remark 2), we obtain the four complex parameter solution Eq.(\ref
{flatGoodCut}) 
\begin{eqnarray}
u &=&\frac{\mu ^{0}+\widetilde{\zeta }\mu ^{1}}{1+\zeta \widetilde{\zeta }}=%
\frac{\alpha +\beta \widetilde{\zeta }+\widetilde{\beta }\zeta +\gamma \zeta 
\widetilde{\zeta }}{1+\zeta \widetilde{\zeta }}=z^{a}l_{a}  \label{FlatCut}
\\
\overline{L} &=&\frac{\mu ^{1}-\zeta \mu ^{0}}{P}=z^{a}\overline{m}_{a}
\label{flatLbar} \\
l_{a} &\equiv &\frac{\sqrt{2}}{2}(1,-\frac{\zeta +\overline{\zeta }}{1+\zeta 
\overline{\zeta }},i\frac{\zeta -\overline{\zeta }}{1+\zeta \overline{\zeta }%
},\frac{1-\zeta \overline{\zeta }}{1+\zeta \overline{\zeta }})  \label{l} \\
\overline{m}_{a} &\equiv &\frac{\sqrt{2}}{2}(0,\frac{1-\zeta ^{2}}{1+\zeta 
\overline{\zeta }},\frac{i(1+\zeta ^{2})}{1+\zeta \overline{\zeta }},\frac{%
2\zeta }{1+\zeta \overline{\zeta }})  \label{mbar}
\end{eqnarray}
We now chose the four parameters $z^{a}\Leftrightarrow (\alpha ,\beta ,%
\widetilde{\beta },\gamma )$ to be arbitrary complex analytic functions of a
`complex-time' parameter, $\tau ,$ i.e., 
\begin{equation}
z^{a}=\xi ^{a}(\tau ),  \label{curve}
\end{equation}
From this it follows that both 
\begin{eqnarray}
\mu ^{0}(\tau ,\zeta ) &=&u-\overline{L}\overline{\zeta }  \label{mu0} \\
\mu ^{1}(\tau ,\zeta ) &=&\overline{L}+\zeta u  \label{mu1}
\end{eqnarray}
are functions \textit{only} of ($\tau ,\zeta $) so that by eliminating $\tau 
$ between them, we define, \textit{implicitly}, a complex analytic function $%
F(\mu ^{0},\mu ^{1},\zeta )=0.$ We refer to this class of functions as
yielding the restricted Kerr theorem. We thus have

\begin{corollary}
Every complex analytic curve, $z^{a}=\xi ^{a}(\tau ),$ on the four-complex
dimensional parameter space of solutions to the flat good-cut equation, Eq.(%
\ref{flatGoodCut}), determines a shear-free null geodesic congruence in
Minkowski space. This is accomplished by the elimination of $\tau $ between
the two equations $u=\xi ^{a}(\tau )l_{a}(\zeta ,\overline{\zeta })$ and $%
\overline{L}=\xi ^{a}(\tau )\overline{m}_{a}(\zeta ,\overline{\zeta }).$
Note that these relations have a simple geometric meaning: for each point on 
$\frak{I}^{+},$ i.e., at ($u$,$\zeta ,\overline{\zeta }$), there is a
direction on its past light-cone given by $L$($u$,$\zeta ,\overline{\zeta }$%
) that yield the initial conditions for a null geodesic going backwards into
the flat space-time producing a shear-free null geodesic congruence.
\end{corollary}

The generalization of this result to asymptotic twistor theory is simple. It
is known\cite{NKI} that complex analytic curves on the solution space to the
good-cut equation, i.e., on H-Space, determine \textit{\ asymptotically
shear free null geodesic congruences} in the neighborhood of $\frak{I}^{+}$
of asymptotically flat space-times. In other words, given a Bondi shear and
the solution 
\begin{equation}
u=Z(z^{a},\zeta ,\widetilde{\zeta })  \label{sol2}
\end{equation}
to the good-cut equation, with the arbitrary choice of the complex
world-line $z^{a}=\xi ^{a}(\tau ),$ the pair 
\begin{eqnarray}
u &=&Z(\xi ^{a}(\tau ),\zeta ,\overline{\zeta })\equiv X(\tau ,\zeta ,%
\overline{\zeta })  \label{pair1} \\
\overline{L} &=&\overline{\text{\dh }}Z(\xi ^{a}(\tau ),\zeta ,\overline{
\zeta })\equiv \overline{L}(\tau ,\zeta ,\overline{\zeta }),  \label{pair2}
\end{eqnarray}
or equivalently from 
\begin{eqnarray}
\mu ^{0}(\tau ,\zeta ,\overline{\zeta }) &=&u-\overline{L}\overline{\zeta }
\label{p1} \\
\mu ^{1}(\tau ,\zeta ,\overline{\zeta }) &=&\overline{L}+\zeta u,  \label{p2}
\end{eqnarray}
determines an asymptotically shear free null geodesic congruence.

\begin{theorem}
The single function, defined implicitly by the elimination of $\tau $ from
this pair, (the \textit{generalized restricted Kerr theorem)} is the
analogue of the restricted Kerr theorem but now it determines \textit{\
asymptotically shear-free congruences}.
\end{theorem}

Again, there is the same simple geometric meaning for this pair as in the
Corollary.

\begin{remark}
In Penrose and Rindler, Volume II\cite{PandR}, there is a rather abstract
discussion of hypersurface twistors and a generalized version of the Kerr
theorem to that situation. They remark that their version can be extended to 
$\frak{I}^{+}.$ It appears almost certain that our version is a restriction
or subset of theirs when applied to $\frak{I}^{+}$.
\end{remark}

\section{$\frak{I}^{+}$ as a three-dimensional CR manifold}

We have seen that given an arbitrary asymptotically flat Einstein or
Einstein-Maxwell space-time, we can define both a projective asymptotic
twistor space and its dual space. From this pair a scalar product can be
defined which allows the restriction to the five dimensional $subspace$ of
projective twistors with a vanishing scalar product, i.e., to the null
twistor space $P\frak{N.}$ Further we saw that this restriction allows a CR
structure to be defined on $P\frak{N}$ which we can identify with the sphere
bundle over $\frak{I}^{+}$ with the local `real' coordinates $(u$,$L,%
\overline{L},\zeta ,\overline{\zeta }$).

By choosing a curve, $z^{a}=\xi ^{a}(\tau )$ in H-Space, our version of the
Kerr theorem allows us, by choosing a cross-section of the sphere-bundle, to
map the five dimensional CR manifold $P\frak{N}$ into $\frak{I}^{+}$. From
this, $\frak{I}^{+}$ inherits three-dimensional CR structures that are
associated with each choice of H-Space curve. To see this all we must do is
choose $\overline{L}$ as the function on $\frak{I}^{+}$ given by the pair,
Eqs.(\ref{pair1}) and (\ref{pair2}), with the $\tau $ eliminated. This
yields, in principle, 
\begin{equation}
\overline{L}=\overline{L}^{*}(u,\zeta ,\overline{\zeta })  \label{L*}
\end{equation}
with

\[
d\overline{L}=\partial _{u}\overline{L}^{*}du+\partial _{\zeta }\overline{L}%
^{*}d\zeta +\partial _{\overline{\zeta }}\overline{L}^{*}d\overline{\zeta }. 
\]
The CR structure of $P\frak{N,}$ i.e., Eqs.(\ref{CR1}),(\ref{CR2}) and (\ref
{CR3}), reduces to

\begin{eqnarray}
l &=&du-\frac{L}{(1+\zeta \overline{\zeta })}d\zeta -\frac{\overline{L}}{
(1+\zeta \overline{\zeta })}d\overline{\zeta }  \label{cr1} \\
\overline{m} &=&\frac{d\zeta }{(1+\zeta \overline{\zeta })}  \label{cr2}
\end{eqnarray}
with ($L,\overline{L}$) given by (\ref{L*}). The third one-form, $N$, now is
a combination of $l$ and $\overline{m}.$ For each choice of the curve, $%
z^{a}=\xi ^{a}(\tau ),$ we have different CR structure on $\frak{I}^{+}.$

In addition, the complex parameter $\tau $ has the meaning of a CR function
and in fact the pair, ($\tau ,\zeta $), form the local $\mathbf{C}^{2}$
whose real three-dimensional subspace is $\frak{I}^{+}.$ We thus have that $%
\frak{I}^{+}$ is a locally embeddable CR manifold. The parametric form of
the mapping is given by ($\tau ,\zeta )=(T(u,\zeta ,\overline{\zeta }),x+iy$%
) with $T(u,\zeta ,\overline{\zeta })$ given by the inversion of the
function Eq.(\ref{pair1}), i.e., from $u=X(\tau ,\zeta ,\overline{\zeta }).$
By taking the exterior derivatives of ($\tau ,\zeta )$ one can easily
reconstruct the forms (\ref{cr1})-(\ref{cr2}). \{See Newman and Nurowski\cite
{NandN} for details.\}

\begin{remark}
Strictly speaking , for the complex equation, $\tau =T(u,\zeta ,\overline{%
\zeta }),$ to make sense, it should first be defined on $\frak{CI}^{+}$ and
then restricted to $\frak{I}^{+}$
\end{remark}

Recently\cite{NKI,NKII,NKO,NO} it was shown, for all asymptotically flat
Einstein space-times, that there is a canonical choice of the H-space curve
and hence a unique CR structure on $\frak{I}^{+}.$ The uniqueness arises
from certain properties of the Weyl tensor. For asymptotically flat
Einstein-Maxwell space-times there are two canonical choices for the curve
and the CR structure. One choice comes from the Weyl tensor's properties,
the other from those of the Maxwell tensor. These unique curves appear to
have a considerable physical meaning\cite{NKO,NO} but a discussion here
would be too far afield.

\section{The Einstein Equations on P$\frak{N.}$}

Though it would be nicer if we could write the asymptotic Einstein
equations, (i.e., the asymptotic Bianchi Identities) as equations solely
involving the variables defining P$\frak{N,}$ i.e., $(u$,$\overline{L},\zeta 
$), at least in the case of asymptotically flat algebraically special vacuum
metrics, we can come close to doing so. Aside from the P$\frak{N}$
variables, introducing only the `Bondi mass aspect', $\Psi (u,\zeta ,%
\overline{\zeta }),$ which is real, we can write the asymptotically flat
algebraically special vacuum Einstein equations\cite{NO,NandT} in the
following fashion:

First defining the Bondi asymptotic shear in terms of ($\overline{L},L$) by,

\begin{equation}
\overline{\sigma }=\overline{\sigma }[\overline{L}]\equiv \overline{\text{%
\dh }}(\overline{L})+\frac{1}{2}(\overline{L}^{2}),_{u}  \label{sigmabar}
\end{equation}
the full Einstein equations are given by,

\begin{eqnarray}
\text{\dh }\Psi &=&\text{\dh }[\overline{\sigma },_{u}\sigma ]+\text{\dh }
^{3}\overline{\sigma }+2\sigma \text{\dh }(\overline{\sigma },_{u})+3L\text{%
\dh }^{2}(\overline{\sigma },_{u})+L^{3}\overline{\sigma },_{uuu}+3L\sigma 
\overline{\sigma },_{uu}+3L^{2}\text{\dh }(\overline{\sigma },_{uu})
\label{GR1} \\
&&-3L,_{u}[\Psi -2L\text{\dh }(\overline{\sigma },_{u})-L^{2}\overline{
\sigma },_{uu}-\text{\dh }^{2}\overline{\sigma }-\sigma \overline{\sigma }%
,_{u}],  \nonumber \\
\Psi ^{\cdot } &=&\sigma ,_{u}\overline{\sigma },_{u}.  \label{GR2}
\end{eqnarray}

For the case of the general asymptotically flat Einstein equations, there is
a generalization of the above but the role of the Bondi mass aspect has to
be considerably generalized to include several other functions that are
given as initial conditions with their evolution determined by the P$\frak{N}
$ variables and the attractiveness is lost.

\section{Conclusion}

The work described here can be considered as generalizing to arbitrary
asymptotically flat Einstein or Einstein-Maxwell space-times the work done
by many authors\cite{Tr,Taf,LN,LNT,LNT2,PNAT,PandR,PM,RP} on structures that
exist on or are associated with \textit{special space-times}, specifically,
Minkowski space and the algebraically special space-times. These structures,
which included twistor and projective twistor spaces with several of their
three and five dimensional real subspaces which come with CR structures,
show the rich mathematical structures contained in large families of
solutions to the Einstein equations. It also underlines the deep connection
that shear-free and asymptotically shear-free null geodesic congruences play
in these structures and in the very basic structure of the Einstein
equations. Though it is not at all clear what are the physical meanings and
consequences of the existence of these structures, there have been strong
suggestions of possible physical significance\cite{NKI,NKII,NKO,NO} that
must be further explored.

\section{Acknowledgments}

This material is based upon work (partially) supported by the National
Science Foundation under Grant No. PHY-0244513. Any opinions, findings, and
conclusions or recommendations expressed in this material are those of the
authors and do not necessarily reflect the views of the National Science.
E.T.N. thanks the NSF for this support. He also thanks both Roger Penrose
and Pawel Nurowski for certain research suggestions and Jerzy Lewandowski
for great help in understanding CR structures.

\end{document}